# Ich weiß, was du nächsten Sommer getan haben wirst: Predictive Policing in Österreich

Angelika Adensamer und Lukas Daniel Klausner

Zusammenfassung. Predictive Policing ist ein datenbasiertes und prognosegetriebenes Modell für Polizeiarbeit. Wir geben in diesem Artikel einen Überblick über den aktuellen Stand in Österreich und diskutieren technische, politisch-gesellschaftliche und rechtliche Probleme, die sich daraus ergeben – etwa das mangelhafte Bewusstsein für Prozesse gesellschaftlicher Diskriminierung, die verzerrte Datenbasis, die PP zugrundeliegt, und fehlende Reflexion über zugrundeliegende Annahmen und Rückkopplungseffekte. Anlasslose Grundrechtseingriffe sind weder durch die StPO noch das SPG oder das PStSG gedeckt; dem Grundgedanken, dass Polizei erst bei konkreter Gefahrenlage oder Tatverdacht tätig werden darf, muss weiterhin Rechnung getragen werden. Aus unserer Sicht sollte angesichts der zahlreichen Probleme (und auch aus rechtspolitischen Erwägungen) auf PP verzichtet werden und stattdessen Ressourcen und Überlegung in die Lösung jener gesellschaftlicher Probleme investiert werden, die zu Kriminalität führen.

Abstract. Predictive policing is a data-based, predictive analytical technique used in law enforcement. In this paper, we give an overview of the current situation in Austria and discuss technical, sociopolitical and legal questions raised by the use of PP, such as the lack of awareness of discriminatory structures in society, the biases in data underlying PP and the lack of reflection on the basic premises and feedback mechanisms of PP. Violations of fundamental rights without cause are not allowed by the Austrian Code of Criminal Procedure (*Strafprozeßordnung*, StPO), the Security Police Act (*Sicherheitspolizeigesetz*, SPG) or the Act concerning Police Protection of the State (*Polizeiliches Staatsschutzgesetz*, PStSG); the principle of allowing police intervention only on the basis of concrete threats or suspicion must remain absolute. Considering the numerous problems (not least from the point of view of legal policy), we conclude that the use of PP should be eschewed and that resources and planning should instead be focussed on solving the social problems which actually cause crime.

## 1. Was ist Predictive Policing?

Predictive Policing (PP) wird ins Deutsche als „prognosebasierte",[1] „vorhersagebasierte"[2] oder „vorausschauende" Polizeiarbeit[3] übersetzt. Die Definition von PP besteht aus zwei Elementen: 1. die polizeiliche Prognose und 2. das dahinterstehende computergestützte Verfahren. Dass etwas vorhergesehen werden muss, um in diesem Sinne „predictive" zu sein, ist klar; was genau vorhergesagt wird, unterscheidet sich aber je nach Software, Modell und Maßnahme. Eine „Vorhersage" kann in diesem Kontext immer nur eine Wahrscheinlichkeitsaussage bedeuten, niemals eine Prognose „mit Sicherheit", so *Belina*.[4] Der Begriff „Predictive Policing" wird dann verwendet, wenn die Prognose auf computergestützten analytisch-technischen Verfahren[5] beruht. Dies sind üblicherweise quantitative,[6] datenbasierte[7] Verfahren.

---

*Schlüsselwörter.* Algorithmen, Big Data, Prediction, Predictive Policing, vorhersagende Polizeiarbeit.

[1] *Egbert*: Predictive Policing, S. 242.
[2] *Egbert*: Siegeszug, S. 17.
[3] *Gerstner*: Predictive Policing, S. 2.
[4] *Belina*: Predictive Policing, S. 88.
[5] *Egbert*: Predictive Policing, S. 242; *Perry* et al.: Predictive Policing, S. 1; *Gerstner*: Predictive Policing, S. 3; *Uchida*: Predictive Policing, S. 3871 ff.
[6] *Perry* et al.: Predictive Policing, S. 1; *Belina*: Predictive Policing, S. 86.
[7] *Gerstner*: Predictive Policing, S. 2; *Gluba*: Predictive Policing – eine Bestandsaufnahme, S. 7.

Über diese Gemeinsamkeiten hinaus bestehen einige Unterschiede zwischen verschiedenen Maßnahmen des PP. Manche gehen mit Eingriffen in subjektive Rechte einher, wenn sie z. B. Personenkontrollen nach sich ziehen, andere werden nur zum Management von Streifendiensten verwendet. Manche operieren auf Basis personenbezogener Daten, andere ohne; manche beziehen die Prognosen auf Orte, wie z. B. das Pilotprojekt in Baden-Württemberg, in dem die Software PRECOBS zur Vorhersage von Orten genutzt wird, an denen Einbruchsdiebstähle am wahrscheinlichsten sind[8] (sog. Hot-Spots, auch als „Risk Terrain Modelling" bezeichnet), andere auf Personen (wie z. B. die Strategic Subject List in Chicago[9]).

Aufgrund der Vielschichtigkeit der Problemstellungen von PP haben wir uns dazu entschieden, in diesem Artikel mit einem interdisziplinären Ansatz problematische Aspekte technischer, praktischer und gesellschaftlicher Natur von PP überblicksmäßig zu umreißen. In rechtlicher Hinsicht wird die Frage erörtert, inwieweit die Polizei heute überhaupt befugt ist, ohne Anfangs- oder Gefahrenverdacht Ermittlungsmaßnahmen zu setzen. Dass das neue PNR-G[10] dies möglich macht, ist ein Novum für die österreichische Rechtsordnung. Schließlich wird auch ein Überblick über in Österreich eingesetzte bzw. entwickelte PP-Modelle gegeben.

## 2. Technische und praktische Probleme von PP

2.1. Garbage In, Garbage Out – Daten und ihre Bewertung. Schon die technischen und praktischen Aspekte von PP bergen viele Tücken. Die erste Problemquelle liegt schon in den Daten selbst:[11] Diese bilden naturgemäß oft die bestehenden Ungleichheiten und Diskriminierungen in der Gesellschaft ab.[12] Wird nicht bewusst und wohlüberlegt versucht, diese Differenzen auszugleichen, reproduziert nur auf solchen Daten basierendes Handeln diese wiederum.[13] Auch die Auswirkungen diskriminierender Polizeiarbeit der Vergangenheit sind so in kriminologischen Daten sichtbar (*dirty data*, analog zu *dirty policing*),[14] nicht zuletzt auch wegen „Under-Reporting" von Sexualstraftaten[15] oder rassistischen Verbrechen.[16] Es wäre die Pflicht der PP-Modelle anwendenden Behörde, dieser Verzerrung in der Datenbasis entgegenzuwirken, berührt ein durch Probabilistik (mit-)geleitetes Agieren der Exekutive doch das Verbot von Diskriminierung durch die Polizei.[17] Eine Voraussetzung hierfür, an der es der Polizei oft mangelt, ist sowohl eine entsprechend diverse Belegschaft als auch ein Bewusstsein für diskriminierende Strukturen;[18] dieser Mangel steht mit der Ungleichbehandlung von Minderheiten durch die Polizei in Wechselwirkung und erzeugt einen zusätzlichen Bias.[19] Eine ähnliche Problematik besteht auch bzgl. der Diversität sowie sozialwissenschaftlicher und ethischer Grundbildung der für das System verantwortlichen Techniker*innen.

---

[8] *Gerstner*: Predictive Policing, S. 1.
[9] S. z. B. *Sheehey*: Algorithmic Paranoia.
[10] PNR-Gesetz BGBl I 2018/64.
[11] In der Informatik spricht man hier gerne von „garbage in, garbage out" und meint damit, dass schlechter oder fehlerhafter Input für ein System ebenso schlechten oder falschen Output dieses Systems bedeutet, egal wie gut das System selbst funktioniert.
[12] *Hao*: Crime-Predicting AIs.
[13] *Singelnstein*: Predictive Policing, S. 4.
[14] „Dirty Policing" steht als Sammelbegriff für verfehlte, diskriminierende und tw. sogar gesetzeswidrige Praktiken und Verhaltensweisen in der Polizeiarbeit; analog dazu wurde der Begriff „Dirty Data" gebildet, s. *Richardson/Schultz/Crawford*: Dirty Data, S. 192.
[15] *Taylor/Gassner*: Stemming the Flow, S. 241 ff.
[16] *Kushnick*: 'Over Policed and Under Protected', ¶1.7.
[17] Vgl. z. B. § 5 Abs 1 Verordnung des Bundesministers für Inneres, mit der Richtlinien für das Einschreiten der Organe des öffentlichen Sicherheitsdienstes erlassen werden (Richtlinien-Verordnung – RLV) BGBl 1993/266 idF BGBl II 2012/155.
[18] *Myers West/Whittaker/Crawford*: Discriminating Systems, S. 15 ff.
[19] *Legewie/Fagan*: Group Threat.

Weiters ist auch die Interpretation der Primärdaten oft problematisch. Oberflächliche Betrachtung der räumlichen Verteilung von Fallhäufigkeiten führt oftmals einfach zu einer Konzentration auf Orte mit großer Bevölkerungsdichte oder mit aus ähnlich trivialen Gründen erhöhtem Datenaufkommen. Auch die Wahl der richtigen Bezugsgröße hat schon einen großen Einfluss auf die Bewertung, wenn man etwa statt der Wohnbevölkerung die tatsächlich anwesenden Personenzahlen (*ambient population*) heranzieht.[20] Selbst die bloße Mustererkennung ist auf vielerlei Arten fehleranfällig: Sie nimmt zwangsläufig eine gewisse Regularität des Verbrechens an (was gerade bei selteneren Straftaten kaum zutrifft), verleitet zur Symptombekämpfung[21] und verschleiert diskriminierende Praktiken in der Polizeiarbeit nach außen wie innen.[22]

2.2. PROBABILISTIK UND MODELLIERUNG. Die Modellierung soziologischer Phänomene durch naturwissenschaftliche oder naturwissenschaftlich motivierte Algorithmen wirft weitere Fragen auf. Oft werden gar keine (wie beim rein korrelativ arbeitenden *HunchLab*[23]) oder vereinfachende Grundannahmen getroffen oder ohne zugrundeliegende soziologische methodologische Überlegungen Algorithmen eingesetzt, die zu wissenschaftlich nicht schlüssig herleitbaren Prognosen führen. In vielen Fällen wird nicht oder bestenfalls am Rande reflektiert, dass allen derartigen algorithmischen Zugängen zahlreiche implizite Grundannahmen zugrunde liegen.[24] In Oakland wurde z. B. ein PP-Algorithmus getestet, der auf Basis der sog. „Near-Repeat"-Kriminalitätstheorie von seismographischen Modellen inspiriert ist,[25] dessen zentrale Berechnung aber auf einen schlichten gleitenden Mittelwert hinausläuft und keinerlei Rückkopplungseffekte auf das Kriminalitätsaufkommen[26] berücksichtigt.[27] Auch dass verschiedene Bevölkerungsgruppen unterschiedliche Elastizitäten in Bezug auf die Verübung von Straftaten aufweisen,[28] wird nicht in Betracht gezogen und kann die erwarteten Ergebnisse ins Gegenteil verkehren.[29] Generell ist eine wichtige Frage für jegliche statistisch angeleitete Strategie, wie das modellgeleitete Agieren wiederum das kriminelle Verhalten beeinflusst und welche Effekte zweiter und höherer Ordnung (d. h. kaskadierende und rückbezügliche Folgeeffekte[30]) das wiederum auf die Exekutivarbeit hat.[31]

2.3. KONTROLLE UND EVALUIERUNG. Schlussendlich bleiben auch in der Beurteilung und Einordnung der algorithmisch getroffenen oder von Algorithmen unterstützten Entscheidungen noch große Lücken. Mangelnde Transparenz macht viele algorithmische Systeme zu reinen Black Boxes[32] und erschwert oder verunmöglicht eine kritische Betrachtung.[33]

---

[20] *Belina*: Predictive Policing, S. 92 ff.
[21] Gemäß *Goodharts* Gesetz: „When a measure becomes a target, it ceases to be a good measure."
[22] *Kaufmann/Egbert/Leese*: Politics of Patterns, S. 11 ff.
[23] *Shapiro*: Reform Predictive Policing, S. 459.
[24] *Bennett Moses/Chan*: Algorithmic Prediction, S. 809 ff.
[25] *Mohler* et al.: Field Trials.
[26] Hiermit ist gemeint, dass die Aktivität der Polizei wiederum Auswirkungen auf das Kriminalitätsaufkommen hat, was sich wiederum auf das Handeln der Polizei auswirkt etc.
[27] *Lum/Isaac*: To Predict and Serve?, S. 18.
[28] „Elastizität" bezeichnet in den Wirtschafts- und Sozialwissenschaften die Beeinflussbarkeit einer gewissen Größe relativ zu einer anderen Größe. Hier ist damit gemeint, dass verschiedene Bevölkerungsgruppen unterschiedlich auf das Poliziertwerden reagieren, d. h. dass sich ihr Verhalten in Abhängigkeit vom Handeln der Polizei in verschiedener Weise bzw. unterschiedlich stark verändert.
[29] *Harcourt*: Against Prediction, S. 23 f.
[30] Siehe Fußnote 26.
[31] *Richardson/Schultz/Crawford*: Dirty Data, S. 20 ff.
[32] Unter „Black Box" versteht man in der Systemtheorie ein (mglw. sehr komplexes) System, dessen innere Funktionsweise nicht bekannt ist und das daher nur auf Basis reiner Input–Output-Korrespondenzen betrachtet und verstanden werden kann, ohne interne Zusammenhänge und Vorgänge zu kennen.
[33] *Ferguson*: Policing Predictive Policing, 1165 ff. Zu diesem Themenkreis gehört auch die Frage der Verantwortlichkeit, siehe folgender Abschnitt.

Datengeleiteten Entscheidungen wird zudem Objektivität und Neutralität zugeschrieben (und dabei nicht zuletzt der Bias in der Datenbasis durch diskriminierende Polizeipraktiken der Vergangenheit verschleiert), wodurch häufig auch die Notwendigkeit für Begründung und Rechtfertigung des eigenen Handelns negiert wird.[34]

Zuletzt fehlt nach wie vor eine kritische Evaluierung der Effektivität sowie der Vor- und Nachteile von PP-Verfahren.[35] Eine rezente groß angelegte Literaturstudie kam zu dem Resultat, dass bislang kaum empirische Belege vorliegen, weder bzgl. des versprochenen Nutzens noch bzgl. der von Expert*innen befürchteten Nachteile.[36] PP-Modelle sind in aller Regel schon deswegen kaum evaluierbar, da nicht erfüllte Prognosen immer sowohl einem Effekt des Einschreitens als auch einem Fehler der Prognose zuschreibbar sind. Die Schwierigkeiten in der Evaluierung lassen auch an der Ernsthaftigkeit der Beteuerungen von Ressourceneinsparung und Effizienzsteigerung zweifeln, die oftmals mit der Einführung von PP einhergehen.

### 3. Politische und gesellschaftliche Fragestellungen

Andere Probleme entstehen nicht erst durch den Einsatz von Algorithmen, werden dadurch aber (teils empfindlich) verschärft; auf diese Aspekte können wir hier aus Platzgründen nur relativ kurz eingehen.[37] Einerseits besteht die Gefahr von Feedbackloops: Über jene Regionen, in die die Polizei öfter geschickt wird, wird auch mehr Datenmaterial gesammelt, was wiederum mehr Anlass geben kann, Streifen dorthin zu entsenden, wodurch es zu einem positiven Regelkreis kommen kann.[38] Andererseits gibt es aber auch (im Zusammenspiel mit der im vorigen Abschnitt erwähnten Problematik der Symptombekämpfung) klassische Verdrängungseffekte, d. h. gewisse Formen von Kriminalität werden durch räumlich fokussierten Polizeieinsatz nur verlagert.[39] Die Frage, die beim Einsatz neuer Technologien oftmals auf der Strecke bleibt, ist, welche Polizeiarbeit überhaupt wie effektiv ist – und nicht nur wo, wann und gegen wen sie eingesetzt wird. Die ausgeschickte Polizeistreife wird gesamtgesellschaftliche Probleme, die die Sozialwissenschaft als kriminogene Faktoren identifiziert hat, wie z. B. gesellschaftlichen Ausschluss und Perspektivenlosigkeit, nicht lösen, selbst wenn sie noch so effizient eingesetzt wird.

Ein weiterer großer Themenkomplex ist die Frage der Verantwortlichkeit, im rechtlichen wie im ethischen Sinne; insbesondere stellt sich die Frage der Teilung der Verantwortlichkeit zwischen Entwickler*innen von PP-Systemen, den Einsatz solcher Systeme beschließenden Führungskräften und tatsächlich für die Letztentscheidungen verantwortlichen Polizist*innen. Auch die Tatsache, dass solche Algorithmen gem. Art 11 der Polizei-Richtlinie[40] bei Entscheidungen zum Nachteil von Personen nur zur Entscheidungsunterstützung eingesetzt werden und die Letztentscheidung von Menschen getroffen wird, kann nur bedingt beruhigen. Die „Vorschläge" des Systems werden in vielen Fällen nicht hinterfragt und fallweise sogar wider besseres Wissen der Fachleute übernommen (*algorithmic bias* bzw. *automation bias*).[41] Diese teilweise oder gar vollständige Entkopplung von Verantwortung wirft bereits jetzt konkrete rechtliche Fragen in Bezug auf die

---

[34] *Lum/Isaac*: To Predict and Serve?, S. 18 f.; *Bennett Moses/Chan*: Algorithmic Prediction, 817 f. Manche Systeme wie *HunchLab* werben sogar explizit damit, rein korrelativ zu arbeiten und somit angeblich unvoreingenommen zu sein; *Shapiro*: Reform Predictive Policing, S. 459.

[35] Vgl. *Belina*: Predictive Policing, S. 93 f. mit weiteren Nachweisen; *Bennett Moses/Chan*: Algorithmic Prediction, S. 815 ff.

[36] *Meijer/Wessels*: Review of Benefits and Drawbacks.

[37] Für einen generellen Überblick siehe z. B. *Kretschmann*: Wuchern der Gefahr; *Singelnstein*: Predictive Policing.

[38] *Ensign* et al.: Runaway Feedback Loops.

[39] *Belina*: Predictive Policing, S. 93; *Singelnstein*: Predictive Policing, S. 4.

[40] RL (EU) 2016/680 des Europäischen Parlaments und des Rates vom 27. 4. 2016 zum Schutz natürlicher Personen bei der Verarbeitung personenbezogener Daten durch die zuständigen Behörden zum Zwecke der Verhütung, Ermittlung, Aufdeckung oder Verfolgung von Straftaten oder der Strafvollstreckung sowie zum freien Datenverkehr und zur Aufhebung des Rahmenbeschlusses 2008/977/JI des Rates, ABl L 2016/119, 89.

[41] *Goddard/Roudsari/Wyatt*: Automation Bias, S. 123 ff.; *Korff/Georges*: Passenger Name Records, S. 29 f.

Rechtsstaatlichkeit des Verfahrens auf.[42] Hier sind noch unzählige Fragen ungeklärt, beginnend bei ganz grundlegenden wie der zurechenbaren Verantwortlichkeit für Entscheidungen, wenn die Letztentscheider*innen weder über den Einsatz von Algorithmen (mit-)bestimmen können noch diese in ausreichendem Umfang verstehen können (sei es mangels technischen Wissens oder wegen fehlender Transparenz der algorithmischen Systeme). Bis auf erste Ansätze[43] steht die Forschung und Politik hier noch ganz am Anfang.

Auch die bereits im vorigen Abschnitt diskutierte Frage der Datengrundlage ist nicht nur eine technische, sondern auch eine politische Frage,[44] insbesondere die Auswahl und Abstraktion der durch die Daten vermeintlich objektivierten Tatbestände und Sachverhalte.[45] Da PP-Modelle gerade darauf aufbauen, lückenlos alle Gebiete bzw. Personen auf kriminogene Faktoren hin zu analysieren und miteinander zu vergleichen, bedeutet ihre Anwendung notgedrungen, dass niemand komplett unbeobachtet bleiben kann. Kombiniert mit der Möglichkeit, schon einzuschreiten, bevor überhaupt Straftaten passiert sind,[46] ergibt sich die Gefahr eines Staates ohne Freiheit vor Überwachung und ohne das Anerkenntnis, dass sich die Gesamtheit einzelner individueller Entscheidungen, auf die es schlussendlich ankommt, in mathematischen Wahrscheinlichkeiten niemals abbilden lassen kann. Den Individuen wird ein generalisierter Verdacht entgegengebracht.[47] Dadurch werden ihre Gruppenzugehörigkeiten konstituiert und hervorgestrichen, während die Bedeutung ihrer Individualität hintangestellt wird. Es ist kein Zufall, dass diese Entwicklungen in Zeiten von nie dagewesenen Datenmengen, höherer und fortgeschrittenerer Analyseleistung sowie einer vorherrschenden (und zumeist wohl politische Interessen verbergenden) Technokratisierung der Politik geschehen.

## 4. PP im Lichte polizeilicher Aufgaben

Aus der Verwendung von Methoden des PP ergeben sich eine Vielzahl wichtiger rechtlicher Fragen: Von der Legalität konkreter Maßnahmen über den Schutz von Grundrechten und vor Diskriminierung, Datenschutz, Rechtsschutz für individuell Betroffene und die rechtliche Qualität algorithmischer Entscheidungen bis hin zur Bedeutung von Algorithmen zur Unterstützung von Entscheidungen auf die Arbeitsverhältnisse von Polizist*innen. Im Folgenden werden wir exemplarisch nur folgende rechtliche Frage anschneiden, ohne sie jedoch abschließend beantworten zu können: Auf Basis welcher gesetzlichen Aufgaben ist die Polizei überhaupt befugt, ohne Anfangs- oder Gefahrenverdacht Ermittlungsmaßnahmen zu setzen? Gibt es keine Aufgaben, zu deren Erfüllung die Polizei PP-Methoden einsetzen kann, ist ihr Einsatz rechtlich unzulässig.

Die Strafprozessordnung bietet eine Rechtsgrundlage für die Aufklärung von Straftaten und die Verfolgung Verdächtiger (§ 1 Abs 1 StPO).[48] Das Strafverfahren beginnt mit dem Anfangsverdacht (§ 1 Abs 2 StPO); der Bezug auf den Verdacht einer Straftat hat laut den Erl. zur StPO-Reform 2004 den rechtsstaatlichen Zweck, Personen davor zu schützen, „ohne gegebenen Anlass zum Objekt von Ermittlungen zu werden"[49] – ein Telos, den der OGH 2012 bestätigte.[50] Schließlich wurde 2014 in § 1 Abs 3 StPO erstmals eine Definition des Anfangsverdachts geschaffen.[51] Dieser liegt erst vor, wenn auf Grund bestimmter Tatsachen angenommen werden kann, dass eine Straftat begangen worden ist. Laut den Erl. soll durch diese Bestimmung sichergestellt werden, dass ein Ermittlungsverfahren erst beginnt, wenn „verifizierbare und widerlegbare Anhaltspunkte" vorliegen.[52] Mangelt es an solchen Anhaltspunkten, „hat die Staatsanwaltschaft von der Einleitung eines Ermittlungsverfahrens abzusehen".[53] Die StPO bietet also keine Rechtsgrundlage für präventive Ermittlungen ohne Anfangsverdacht auf eine begangene Straftat. Folglich scheinen die Entwicklung und der Einsatz von PP-Programmen durch das BKA (s. u.) in Anbetracht dessen, dass dem BKA gem. § 5 Abs 1 u Abs 2 BKA-G[54] ausschließlich kriminalpolizeiliche Aufgaben iSd StPO obliegen, auf rechtlich mehr als wackeligen Beinen zu stehen.

---

[42] S. hierzu etwa *Citron*: Technological Due Process; *van der Sloot/Broeders/Schrijvers* (Hrsg.): Exploring the Boundaries, S. 166 f.
[43] *Citron*: Technological Due Process, S. 1294 ff.; *van der Sloot/Broeders/Schrijvers* (Hrsg.): Exploring the Boundaries.
[44] Für eine plakative Darstellung der Lage in Österreich sowohl hinsichtlich der Datenqualität als auch des Umgangs mit Komplexität seitens des Führungspersonals siehe *Marouschek*: Vom Informationsfriedhof zu Führungsinformationssystemen.
[45] *Belina*: Predictive Policing, S. 94 f.
[46] Wie z. B. im Bereich der erweiterten Gefahrenerforschung in § 6 PStSG, siehe genauer unten.
[47] *Ferguson*: Big Data Policing, S. 126.
[48] Strafprozeßordnung BGBl 1975/631 idF BGBl I 2018/70.
[49] ErläutRV 25 BlgNR 22. GP 26.
[50] OGH 11.6.2012, 1 Präs. 2690–2113/12i.
[51] BGBl I 2014/74.

Während die StPO die Bekämpfung und Aufklärung von Straftaten regelt, sind in SPG und PStSG[55] sicherheitspolizeiliche Aufgaben festgelegt. Diese umfassen die Gefahrenabwehr und die Aufrechterhaltung der öffentlichen Ordnung und Sicherheit (§ 3 SPG iVm Art 10 Abs 1 Z 7 B-VG). Die polizeilichen Aufgaben in SPG und PStSG wurden in den letzten Jahren immer weiter ins Vorfeld verlegt.[56] Die am weitesten ins Vorfeld reichenden Aufgaben, die polizeiliche Ermittlungsmaßnahmen begründen können, sind die (erweiterte) Gefahrenerforschung und die Abwehr allgemeiner Gefahren und wahrscheinlicher Angriffe. Die Gefahrenerforschung ist gem. § 16 Abs 4 SPG[57] die Feststellung einer Gefahrenquelle und des Sachverhalts, soweit dies für die Abwehr einer Gefahr notwendig ist, und obliegt gem. § 28a Abs 1 SPG den Sicherheitsbehörden. Gem. § 21 Abs 1 SPG obliegt den Sicherheitsbehörden auch die Abwehr allgemeiner Gefahren. Diese sind gem. § 16 Abs 1 SPG einerseits gefährliche Angriffe (Z 1 iVm Abs 2 u Abs 3 leg cit) und andererseits kriminelle Verbindungen, also die Verbindung von mindestens drei Personen mit dem Vorsatz, fortgesetzt gerichtlich strafbare Handlungen zu begehen. Die Gefahrenabwehr ist eine Befugnis, die weit ins Vorfeld tatsächlicher Angriffe und Straftaten hineinreicht.[58] Schließlich haben die Sicherheitsbehörden nach § 22 Abs 2 SPG auch die Aufgabe, wahrscheinlichen gefährlichen Angriffen vorzubeugen. Auch wenn diese Aufgaben nicht erst bestehen, wenn eine akute Gefahr besteht, so hängen sie doch von einer „besonderen Situation, in der sich eine akute Gefahr leichter als im Normalfall entwickeln kann", ab.[59] Obwohl diese Aufgaben also sehr weit im Vorfeld liegen, sind sie dennoch nicht komplett losgelöst von konkreten Anlässen und Situationen – wären sie es, wären die Definitionen in § 16 SPG obsolet.

Die erweiterte Gefahrenerforschung wurde 2016 vom SPG in den § 6 des damals neuen PStSG verschoben.[60] Auch hier ist aber der Einsatz von Ermittlungsmaßnahmen nur unter bestimmten Voraussetzungen erlaubt: Gem. § 6 Abs 1 Z 1 können im Zuge der erweiterten Gefahrenerforschung nur Gruppierungen beobachtet werden, in deren Umfeld mit bestimmten Gefahren und Angriffen zu rechnen ist, und gem. § 6 Abs 1 Z 2 besteht die Aufgabe, auf einen begründeten Gefahrenverdacht hin verfassungsgefährdende Angriffe iSd § 6 Abs 2 abzuwehren. *Heißl* erachtet hierfür eine konkrete Bedrohungssituation als notwendig.[61] Ein begründeter Gefahrenverdacht ist dabei laut Erl. jedenfalls mehr als die bloße Möglichkeit oder Nichtausschließbarkeit eines Angriffes.[62] Obgleich auch die Aufgaben nach dem PStSG als zu breit und unbestimmt kritisiert wurden,[63] sind sie nicht völlig unabhängig vom Vorliegen bestimmter Anhaltspunkte für einen Verdacht und können demnach keine Rechtsgrundlage für PP ohne Anlass bieten.

---

[52] 181 BlgNR 25. GP 2.
[53] 181 BlgNR 25. GP 2.
[54] Bundeskriminalamt-Gesetz BGBl I 2002/22 idF BGBl I 2016/118.
[55] Polizeiliches Staatsschutzgesetz BGBl I 2016/5 idF BGBl I 2018/32.
[56] Vgl. dazu die Debatte um die erweiterte Gefahrenerforschung u. a. bei *Kretschmann*: Wuchern der Gefahr.
[57] Sicherheitspolizeigesetz BGBl 1994/505 idF BGBl I 2018/56.
[58] Vgl. z. B. *Zerbes*: Spitzeln, Spähen, Spionieren, S. 259 ff.
[59] *Zerbes*: Spitzeln, Spähen, Spionieren, S. 262.
[60] BGBl I 2016/5.
[61] *Heißl*: Polizeiliches Staatsschutzgesetz, § 6, Rz 23.
[62] ErläutRV 763 BlgNR 25. GP 4 mVa *Hauer/Keplinger*: Sicherheitspolizeigesetz, § 22.
[63] *Adensamer/Sagmeister*: Das Polizeiliche Staatsschutzgesetz, S. 67 ff.

Polizeiliche Eingriffe in subjektive Grund- und Freiheitsrechte ohne Gefahren- oder Tatverdacht – also ohne Anlass – sind mit der Rechtsordnung grundsätzlich nicht vereinbar.[64] So schreibt *Zerbes*: „Selbst zur Verteidigung der strafrechtsbewehrten Rechtsgüter greift die öffentliche Gewalt daher nur aus bestimmten Anlässen ein. Durch exakte Definitionen dieser Anlässe wird sie beschränkt."[65] In einem Rechtsstaat, der sich den Freiheitsrechten verpflichtet, müsse ein „staatsfreier Raum" geschaffen werden.[66]

4.1. PASSENGER NAME RECORDS (PNR). Eingriffsschranken für staatliche Eingriffe waren bisher in allen einfachen Gesetzen vorgesehen, die die polizeiliche Tätigkeiten regeln, wie in der StPO, dem SPG und dem PStSG, nicht aber im neuen PNR-G. Die EU-RL 2016/681 über die Verwendung von Fluggastdatensätzen (PNR-Daten) zur Verhütung, Aufdeckung, Ermittlung und Verfolgung von terroristischen Straftaten und schwerer Kriminalität[67] verpflichtet alle Mitgliedstaaten, eine große Menge an Fluggastdaten über alle Personen und Flüge in die und aus der EU zu sammeln, und ermächtigt zur anlasslosen Vorratsdatenanalyse. In Österreich wurde die RL durch das PNR-G umgesetzt, welches seit 2018 in Kraft ist. Das PNR-G stellt also insofern einen Paradigmenwechsel dar, als es Ermittlungshandlungen ohne Anlass erlaubt, was für die österreichische Rechtsordnung ein Novum ist.[68]

Gem. § 2 Abs 1 PNR-G sind Luftfahrtunternehmen verpflichtet, Fluggastdaten iSd § 3 PNR-G vor dem Abflug und nach dem Check-In an die beim Bundeskriminalamt (BKA) eingerichtete Fluggastdatenzentralstelle (Passenger Information Unit, PIU) (§ 1 Abs 2 PNR-G) weiterzuleiten. Die PIU wird durch § 4 Abs 1 ermächtigt, die bei ihr einlangenden Daten mit Fahndungsevidenzen und sonstigen sicherheitspolizeilichen Datenbanken mit dem Zweck der Verbrechensbekämpfung abzugleichen (Z 1) sowie diese „anhand festgelegter Kriterien" zu analysieren (Z 2). Für die Vorratsdatenanalyse nach § 4 Abs 1 Z 2 gibt es keine weiteren gesetzlichen Voraussetzungen, sie kann also verdachts- und anlasslos eingesetzt werden. Somit fällt die Verarbeitung von Fluggastdaten unter die oben stehende Definition von PP. Nach § 4 Abs 1 PNR-G können zu den sehr breit angelegten Zwecken des § 1 Abs 1 PNR-G (also zur Vorbeugung, Verhinderung und Aufklärung bestimmter Straftaten) alle bei der PIU eingelangten Daten sowohl gem. Z 1 mit anderen Datenbanken als auch gem. Z 2 anhand bestimmter „festgelegter Kriterien" iSd § 5 PNR-G abgeglichen werden. Eine engere Einschränkung auf einen Gefahren- oder Tatverdacht gibt es im PNR-G nicht. Dass die anlasslose Verarbeitung von Fluggastdaten einen Eingriff in – und in weiterer Folge auch eine Verletzung von – Art 7 und 8 GRC darstellt, hat schon der EuGH in seinem Gutachten über das Abkommen über den Austausch von PNR-Daten zwischen der EU und Kanada festgestellt.[69] Auch die PNR-RL ist demnach EU-rechtswidrig, Verfahren dagegen wurden schon in Deutschland und Österreich angestrengt.[70]

Der Schutz davor, „ohne gegebenen Anlass zum Objekt von Ermittlungen zu werden",[71] wird durch das PNR-G abgeschafft. Das ist mit rechtsstaatlichen Prinzipien und Freiheitsrechten unvereinbar.[72] Derartig unbegrenzt einsetzbare Befugnisse sind für das österreichische Rechtssystem vollkommen neu. Auch das Rechtsschutzsystem, das in Sicherheitsverwaltung und Kriminalpolizei aufgeteilt ist, wird durch das PNR-G, das zwar polizeiliches Arbeiten regelt, aber keinem von beiden eindeutig zuordenbar ist, auf die Probe gestellt.

---

[64] Vgl. dazu insb. *Zerbes*: Spitzeln, Spähen, Spionieren, S. 242 ff.
[65] *Zerbes*: Spitzeln, Spähen, Spionieren, S. 243.
[66] Vgl. *Zerbes*: Spitzeln, Spähen, Spionieren, S. 254 f.
[67] Abl L 2016/119, 132.
[68] Ebenso für den dt. Rechtsraum, vgl. *Singelnstein*: Predictive Policing, S. 5.
[69] Siehe EuGH 26. 7. 2015 – Gutachten 1/15 Rz 121 ff.
[70] *No PNR*: Wir klagen gegen die massenhafte Verarbeitung von Fluggastdaten!; *epicenter.works*: Wir kippen die Fluggastdatenspeicherung.
[71] ErläutRV 25 BlgNR 22. GP 26.
[72] Vgl. auch *Korff/Georges*: Passenger Name Records, S. 90.

4.2. Andere PP-Maßnahmen. Näherer Betrachtung bedarf daneben auch die Befugnis zum Streifendienst, insb. im Hinblick auf PP-Maßnahmen, die der Polizei eine Entscheidungshilfe zur Hand geben sollen, wo der Streifendienst ausgeführt werden soll. Der einfache Streifendienst iSd § 5 Abs 3 SPG geht ohne Eingriff in subjektive Rechte einher und ist als solches gem. § 28a Abs 2 SPG grundsätzlich zur Aufgabenerfüllung allgemein zulässig.[73] Laut den Erl. zum SPG dient der Streifendienst der „Präventivwirkung durch bloße Anwesenheit" und entspricht einer erhöhten Bereitschaftshaltung, „um bei einer sich konkretisierenden Gefahr schneller einschreiten zu können".[74]

Der Einsatz von PP-Programmen auf der Basis von § 28a Abs 2 SPG ist jedoch – auch wenn je nach konkreter Ausgestaltung rechtlich mglw. zulässig – aus rechtspolitischer Sicht problematisch. Es sollte vermieden werden, durch den Einsatz neuer Software Befugnisse maßgeblich zu erweitern. Insbesondere in Anbetracht oben genannter gesellschaftlicher Gefahren bedarf es politischer Entscheidungen der demokratischen Gesetzgebung, nicht die Schaffung von Gegebenheiten durch technische Anschaffungen und Innovationen seitens der Polizei.

## 5. Predictive Policing in Österreich

Im BKA wird laut einer parlamentarischen Anfragebeantwortung seit 2007 an PP-Modellen gearbeitet,[75] laut *Huberty*, dem Leiter des Büros für räumliche Kriminalanalyse und Geographic Profiling in der Abteilung Kriminalanalyse des BKA, jedoch sogar schon seit 2004.[76] Diese Abteilung, die 2003 ihre Tätigkeit aufnahm, hatte von Beginn an die „zukunftsorientierte" Analyse als Aufgabe.[77] Die Anforderungen an den dort entwickelten Sicherheitsmonitor (SIMO), der 2004 in Betrieb ging[78] und seit 2006 in § 58a SPG gesetzlich verankert ist, inkludierten eine „hohe Qualität der Entscheidungsgrundlagen für Steuerung, Koordinierung und Leitung repressiver und präventiver Maßnahmen, wie Festnahmen, Aufklärungsmaßnahmen der kriminalpolizeilichen Beratungsdienste" etc. sowie die Anwendbarkeit „für wirtschaftlich sinnvolle Ressourceneinsätze bei der Streifenplanung", so der erste Leiter der Kriminalanalyseabteilung des BKA *Marouschek*.[79] Auf Basis der Daten des SIMOs wurde 2007 gemeinsam mit Joanneum Research (JR) das Trend-Monitoring-System (TMS) entwickelt, das für bestimmte Kriminalitätskategorien Wochenprognosen liefert.[80]

Im Förderzeitraum 2009–2011 wurde vom Austrian Institute of Technology das KIRAS-geförderte[81] Projekt *SIS4You* durchgeführt, das eine vom BKA betriebene Plattform zur Einbruchsprävention mit Hilfe der Bevölkerung entwickeln sollte.[82] Ein weiteres KIRAS-gefördertes PP-Projekt von JR war „BASE of ACE: Austrian Crime Explorer", das eine Basis für regionale Kriminalitätsprognosen schaffen sollte und 2012 abgeschlossen wurde.[83]

2013 startete das KIRAS-Projekt Crime Predictive Analytics (*CriPA*).[84] Dieses „widmet sich der vorausschauenden Analyse der Kriminalität in Form von echtzeitfähigen Prognosen und umfasst die Entwicklung geeigneter Algorithmen, Methoden und Softwarekomponenten[,] mit deren Hilfe sich aussagekräftige Modelle und Muster in Kriminalitätsdatenbeständen identifizieren lassen, um auf diese Weise zukünftige Kriminalitätsentwicklungen vorherzusagen oder das Risiko für Straftaten abzuschätzen."[85] Beteiligt waren an diesem Projekt das BKA, JR, das Institut für Rechts- und Kriminalsoziologie und das Z-GIS der Universität Salzburg.[86] Ziel von CriPA war laut Medienberichten ein geographisches Mapping, um Polizeiressourcen gezielter einzusetzen und durch Polizeipräsenz Einbrüche zu verhindern, nach dem Vorbild des Programms *Pre Crime Observation System* (PRECOBS), das in Deutschland und der Schweiz getestet wurde. Das Projekt CriPA wurde 2015 abgeschlossen[87] und sollte ab dem Frühjahr 2015 im Testgebiet Wien eingesetzt werden.[88] Das Modell wurde schließlich von der Polizei nicht übernommen, die Ergebnisse des Projekts flossen aber in ihre Arbeit ein, berichtet Projektleiterin *Kleb*.[89]

---

[73] *Giese*: Sicherheitspolizeirecht, S. 75; *Keplinger/Pühringer*: Sicherheitspolizeigesetz, S. 100.
[74] ErläutRV 148 BlgNR 18. GP 33.
[75] *Sobotka*: Software zur Einbruchsprävention (12318/AB).
[76] Zitiert in *Heitmüller*: Vorzeitige Verhaftung.
[77] Sicherheitsbericht 2003, S. 271.
[78] Sicherheitsbericht 2003, S. 271.
[79] *Marouschek*: Vom Informationsfriedhof zu Führungsinformationssystemen, S. 19.
[80] *Marouschek*: Das Geografische Informationssystem, S. 93.
[81] KIRAS ist das österreichische Förderprogramm für Sicherheitsforschung, vgl. kiras.at.
[82] KIRAS-Projekte 2009–2011, S. 18.
[83] Sicherheitsbericht 2012, S. 225; KIRAS-Projekte 2009–2011, S. 43.
[84] Sicherheitsbericht 2013, S. 64.

2015 berichtete das BMI auch über das Kriminalitätsprognosemodell (KPM).[90] Das Ziel dieser neuen PP-Modelle war laut BMI, „den polizeilichen Personaleinsatz und die Personalkoordination anhand von kurzfristigen räumlich-statistischen Vorhersagemodellen ressourcenschonend und Erfolg versprechend zu verbessern."[91] Auch die Sicherheitsdoktrin des BMI für 2017–2020 formuliert dieses Vorhaben: Neue Technologien wie Big Data oder Artificial Intelligence seien „revolutionäre Ansätze zur Vorbeugung und Bekämpfung der Kriminalität" und das BMI solle sich hier als „First Mover" positionieren.[92] Dies betreffe insbesondere die Nutzung von Videomaterial, aber auch von Kennzeichenerkennungssystemen.[93] 2017 hieß es in einer parlamentarischen Anfragebeantwortung, dass bislang nur eine PP-Methode im Bereich der Einbruchsprävention im Einsatz sei[94] – unklar bleibt aber, auf welcher Rechtsgrundlage.

Ab 2019 solle laut *Huberty* im BKA die Methode des Risk Terrain Modellings (RTM) eingesetzt werden.[95] Hier sollen mittels des geographischen Informationssystems (GIS) nicht personenbezogene Umgebungsfaktoren analysiert werden, wie die Nähe von Parks und Bars oder Altbaubebauung, aber auch die Durchschnittseinkommen nach Wohnort gemäß den Daten der Statistik Austria.[96] Die meisten dieser Daten seien öffentlich verfügbar, so *Huberty*.[97]

Diese PP-Methoden berühren weder den Schutzbereich des Rechts auf Achtung der Privatsphäre (Art 8 EMRK, Art 7 GRC) noch den des Grundrechts auf Datenschutz (Art 1 § 1 Abs 1 DSG,[98] Art 8 GRC), da sie nicht mit der Verwendung von personenbezogenen Daten einhergehen, sondern v. a. der Planung des Streifendienstes dienen. Anders ist es bei dem folgenden Projekt:

Auch in das umstrittene Projekt *INDECT* war mit der FH Technikum Wien eine österreichische Hochschule involviert,[99] welches während seiner Laufzeit 2009–2014 durch die Europäische Kommission mit 10,9 Millionen Euro gefördert wurde.[100] In diesem Projekt sollten personenbezogene Daten aus sozialen Medien mit Vorratsdaten und Videoaufnahmen kombiniert werden, um „abnormales Verhalten" frühzeitig erkennen zu können.[101] Berichte über *INDECT* führten zu internationalen Protesten der Zivilgesellschaft.[102] Weder auf der offiziellen Projekt-Webseite noch

---

[85] Sicherheitsbericht 2013, S. 65.
[86] Sicherheitsbericht 2013, S. 65.
[87] Sicherheitsbericht 2015, S. 84.
[88] *Tempfer*: Haltet den Dieb, bevor er zuschlägt; *Macura*: Pre-Crime-System blickt in die Zukunft.
[89] *Heitmüller*: Vorzeitige Verhaftung.
[90] Sicherheitsbericht 2015, S. 84.
[91] Sicherheitsbericht 2016, S. 82.
[92] Sicherheitsdoktrin des BMI 2017–2020, S. 14.
[93] Sicherheitsdoktrin des BMI 2017–2020, S. 14.
[94] Vermutlich mit Bezug auf CriPA; *Sobotka*: Software zur Einbruchsprävention (12318/AB).
[95] *Heitmüller*: Vorzeitige Verhaftung.
[96] *Al-Youssef*: Predictive Policing.
[97] Zitiert in *Heitmüller*: Vorzeitige Verhaftung.
[98] Datenschutzgesetz BGBl I 1999/165 idF BGBl I 2019/14
[99] *Laub*: INDECT: Anonymous macht gegen totale Überwachung mobil.
[100] *Tajani*: Projekt Indect, Datenschutzverletzung (E-1332/2010 und E-1385/2010).
[101] *Laub*: INDECT: Anonymous macht gegen totale Überwachung mobil.
[102] *Laub*: INDECT: Anonymous macht gegen totale Überwachung mobil.

auf der Seite der FH Technikum Wien sind derzeit weitere Informationen zu diesem Projekt zu finden.

## 6. Fazit

Bestrebungen, Polizeiarbeit verstärkt auf computergestützte Prognosen zu stützen, gibt es in Österreich schon länger, aufgrund der zunehmenden Bedeutung von Big Data und Artificial Intelligence ist mit einer Zunahme und Ausweitung von PP-Methoden in den nächsten Jahren zu rechnen. Die meisten in Österreich entwickelten oder angewendeten PP-Methoden berühren weder den Schutzbereich des Rechts auf Achtung der Privatsphäre (Art 8 EMRK, Art 7 GRC) noch den des Grundrechts auf Datenschutz (Art 1 § 1 Abs 1 DSG, Art 8 GRC), und sollen insb. der Unterstützung des Streifendienstes und der Einbruchsprävention dienen.

Schon angesichts des oft mangelhaften Bewusstseins für Prozesse gesellschaftlicher Diskriminierung und für die verzerrte Datenbasis, die PP zugrunde liegt, ist aus unserer Sicht von einem Einsatz von PP abzusehen. Auch werden die Grundannahmen nicht reflektiert, die hinter den PP-Modellen stehen, und die Effekte nicht mitbedacht, die die Anwendung der Modelle wiederum hat. Jedenfalls sind die bislang durchgeführten Pilotprojekte ausführlich und kritisch zu evaluieren, was bislang nicht erfolgt ist – da wie in Unterabschnitt 2.3 dargelegt eine Evaluierung nach dem aktuellen Stand der Forschung aber prinzipiell schwierig erscheint, müssen hier entweder neue Evaluierungsansätze entwickelt oder, wenn sich auch dies als nicht machbar herausstellt, das Konzept PP als Ganzes infrage gestellt werden. Bestehende Befugnisse dürfen nicht ohne gesetzliche Grundlage durch den Einsatz neuer Software ausgeweitet werden. Nicht zuletzt sollte man dem Einsatz moderner Technik nicht automatisch höhere „Objektivität" zuschreiben – durch den Einsatz von Technologie werden sowohl politische Interessen verschleiert als auch Verantwortung für polizeiliche Entscheidungen abgegeben.

Für die österreichische Rechtsordnung völlig neu und im Hinblick auf den Rechtsschutz problematisch ist das PNR-G, durch das die Analyse von personenbezogenen Fluggastdaten ganz ohne Anlass möglich wird, welche auch einen Grundrechtseingriff in das Recht auf Achtung der Privatsphäre (Art 8 EMRK, Art 7 GRC) und das Grundrecht auf Datenschutz (Art 1 § 1 Abs 1 DSG, Art 8 GRC) darstellt. Anlasslose Grundrechtseingriffe sind weder nach der StPO noch nach dem SPG oder der PStSG rechtmäßig. Dem Grundgedanken, dass die Polizei erst bei konkreten Gefahrenlagen oder einem konkreten Tatverdacht tätig werden darf, muss weiterhin Rechnung getragen werden, um einen Raum der Freiheit vor dem Staat garantieren zu können.

Angesichts der zahlreichen Probleme und offenen Fragen sollte eine freie Gesellschaft auf prognosebasierte Polizeiarbeit verzichten und stattdessen Ressourcen und Überlegung in die Lösung jener gesellschaftlichen Probleme investieren, die zu Kriminalität führen.

## Literatur

Institut für IT-Sicherheitsforschung, FH St. Pölten, Matthias-Corvinus-Straße 15, 3100 St. Pölten, Österreich

*E-Mail*: `mail@l17r.eu`

*URL*: `https://l17r.eu`